\documentclass[10pt]{JHEP3}
\usepackage{epsfig,multicol,bbm}
\usepackage{epsfig}
\usepackage{amsmath}
\usepackage{epsfig}
\usepackage{psfrag}
\usepackage{amsfonts}
\usepackage{graphicx}% Include figure files
\usepackage{dcolumn}% Align table columns on decimal point
\usepackage{bm}% bold math
\usepackage{lineno}% line numbers

\newcommand{\vect}{\left ( \begin{array}{c}}
\newcommand{\evect}{\end{array} \right )}

\usepackage{epsfig}
\usepackage{psfrag}
\usepackage{graphicx}% Include figure files
\usepackage{dcolumn}% Align table columns on decimal point
\usepackage{bm}% bold math
\def\fsl#1{\setbox0=\hbox{$#1$}                 % set a box for #1
   \dimen0=\wd0                                 % and get its size
   \setbox1=\hbox{/} \dimen1=\wd1               % get size of /
   \ifdim\dimen0>\dimen1                        % #1 is bigger
      \rlap{\hbox to \dimen0{\hfil/\hfil}}      % so center / in box
      #1                                        % and print #1
   \else                                        % / is bigger
      \rlap{\hbox to \dimen1{\hfil$#1$\hfil}}   % so center #1
      /                                         % and print /
   \fi}                                         %

\title{The momentum analyticity of two point correlators from perturbation theory and AdS/CFT}

\author{De-fu Hou$^{\dagger a}$, Jia-rong Li$^{\dagger a}$,
Hui Liu $^{\dagger b}$ and Hai-cang Ren $^{\dagger c}$,$^{\dagger a}$\\

{$^{\dagger a}$ Institute of Particle Physics, Huazhong Normal
University, Wuhan 430079, China}\\
{$^{\dagger b}$Physics Department, Jinan University, Guangzhou, China}\\

{$^{\dagger c}$Physics Department, The Rockefeller University,
1230 York Avenue, New York, NY 10021-6399} \\
 }

\abstract {The momentum plane analyticity of two point function of a
relativistic thermal field theory at zero chemical potential is
explored. A general principle regarding the location of the
singularities is extracted. In the case of the $N=4$ supersymmetric
Yang-Mills theory at large $N_c$, a qualitative change in the nature
of the singularity (branch points versus simple poles) from the weak
coupling regime to the strong coupling regime is observed with the
aid of the AdS/CFT  correspondence. }

\begin{document}
%\linenumbers %begin line numbers

%\begin{document}
%%%%%%%%%%%%%%%%%%%%%%%%%%%%%%%%%%%%%%%%
\section{Introduction}
Analyticity of the Green's functions with respect to energy and
momentum is an important property of a quantum field theory.
Extensive investigations have been made mostly on a complex energy
plane  either  at zero temperature\cite{Itzykson} or at nonzero
temperature\cite{Silin,Weldon, pisarski}. The location of the
singularities is dictated by the unitarity and the causality of the
underlying field theory and the nature of the singularities reflects
the character of the excitation spectrum with poles associated to
bound states and the branch cuts to the continuum spectrum of
asymptotic states. By means of the techniques of complex analysis,
general relation between different physical observables (such as
Kramers-Kronig relation) can be extracted without resorting to a
perturbative expansion.

The analyticity of a Green's function with respect to momentum variables is relevant
to the spatial correlation of different operators and static inter-particle potentials\cite{Kapusta}. It
is less explored in literature, probably because of the complication of vector
character of momenta and the lack of general guidelines. For a two point Green's
function of a relativistic field theory, however, simple properties
may be deduced even at a nonzero temperature and this is the main issue addressed in this report.
Our results are two folds:
One concerns the location of the singularity on the complex momentum plane of a two point
Green's function at a nonzero temperature but zero chemical potential.
Our statement applies to any relativistic field theory.
The other is about the nature of the singularities in the weak coupling limit and in the
strong coupling limit and we focus on $N=4$ super-symmetric Yang-Mills theory for
which both limits are under control.

For a relativistic field theory at zero temperature and chemical
potential, a two point Green's function at four-momentum
$Q\equiv(iE,\vec q)$ depends linearly on a number of scalar form
factors which are nontrivial functions of the four-momentum square
$\vec{q}^2-E^2$. Therefore, the analyticity with respect to momentum
follows from that with respect to energy. This is no longer the case
at a nonzero temperature because of the lack of the Lorentz
invariance and the momentum analyticity becomes a separate issue. On
the other hand, the partition function of a relativistic field
theory at a nonzero temperature $T$ can be expressed in terms of a
path integral with the same Lagrangian density as zero temperature
formulated in Euclidean space-time $S^1\times R^3$ with the radius
of $S^1$ equal to $1/T$ so the non-invariance is only imposed by the
boundary condition. Because the $O(4)$ invariance (Euclidean
continuation of the original Lorentz invariance) of the Lagrangian
density, which dimension is associated to the Euclidean time makes
no difference mathematically. Identifying $S^1$ with the Euclidean
time leads to the Matsubara formulation of the thermal field theory
while identifying one of $R^3$ dimensions with the Euclidean time
would end up with a field theory with the same Lagrangian density
but formulated in $S^1\times R^2$ at zero temperature. Consequently,
we can map the momentum plane of the former to the energy plane to
the latter and infer the momentum analyticity of the former from the
unitarity and causality of the later. This way, we deduce that the
two point function with a Euclidean energy is analytic throughout
the complex momentum plane except the imaginary axis for any thermal
relativistic field theory at zero chemical potential. Perturbatively
there is a branch cut running along the imaginary axis, reflecting
the continuum spectrum of the asymptotic states in the interaction
representation of the field theory in $S^1\times R^2$.

Perturbation series of a thermal Yang-Mills theory of three space
dimensions suffers from severe infrared divergence. Even with the
hard-thermal-loop resummation, the perturbation series ceases to
approximate beyond certain orders of the coupling constant $g$. For
the high temperature QCD, the perturbative expansion of the pressure
cannot go beyond $g^6$ and that of a static two point function fails
for small momentum no matter how weak the coupling $g$
is\cite{linde}. The nonperturbative effect behind the infrared
problems may also impact on the analyticity of Green's functions
with respect to energy and momentum. The advent of AdS/CFT duality
enable us to explore the $N=4$ supersymmetric Yang-Mills theory at
large $N_c$ and large 't Hooft
coupling\cite{Maldacena:1997re,Witten:1998qj,MaldacenaReview,SonReview}.
It would be interesting to examine the energy-momentum analyticity
of its Green's functions at nonzero temperature and to compare with
the perturbative results. Such a project has be carried out in the
literature for the energy variable of two point
functions\cite{Starinets,Kovtun}. It was found that, the singularity
on the energy plane remains a cut along the real axis. The analytic
continuation through the cut hits a set of poles associated to the
quasi-normal modes of a black hole. We shall supplement their result
with a study of the analyticity with respect to momentum. Although
the differential equation satisfied by the two point Green's
function are of Heun type which cannot be solved explicitly, the
polynomial dependence of the coefficients of its power series
solution on momentum enable us to make rigorous statements regarding
the analyticity with the aid of Weierstrass theorem. Here we
witnesses a qualitative change in the nature of the singularities
along the imaginary axis: The branch cut in the weak coupling
evolves to a set of simple poles in the strong coupling. To pinpoint
the mechanism of this transition may be highly nontrivial. In this
paper, we merely report our discovery without offering deeper
insights.

In the next section, we shall start with the momentum analyticity of a simple example
of a one-loop self-energy and generalize it to arbitrary number of loops. Then, we shall
extract the correspondence between the thermal field theory in $R^3$ and the zero
temperature field theory in $S^1\times R^2$. The strong coupling limit of the analyticity
will be examined in section III via AdS/CFT duality. In the last section, we discuss our results
together with miscellaneous implications and generalizations. We also speculate the relation of
the poles we discovered to the confinement in two space dimensions. Without special
decleration, the signature of our metric is always Euclidean and we write the four-momentum
of a two point function as $Q=(q_0,\vec q)$ with $\vec q=(0,0,q)$.

%%%%%%%%%%%%%%%%%%%%%%%%%%%%%%%%%%%%%%%%

\section{The two point Green's function of perturbation theory}

%%%%%%%%%%%%%%%%%%%%%%%%%%%%%%%%%%%%%%%%

In this section, we shall demonstrate the momentum plane analyticity of a two point
Green's function from perspective of perturbation theory. We take the
$R$-charge correlator of the $N=4$ SYM as an example in order to compare with the
nonperturbative results at strong coupling in the next section. In the fundamental
representation of the $R$-symmetry group, $SU(4)$, the $R$-charge operator
corresponds to the one of the diagonal generators and we will focus on the
generator
\begin{equation}
Q_R={\rm diagonal}\left(\frac{1}{2},-\frac{1}{2},0,0\right),
\label{rcharge}
\end{equation}
following Ref.\cite{Huot}, which assigns nonzero charges $\pm\frac{1}{2}$ to Weyl fermion fields of the theory. It follows
from the antisymmetric representation of (\ref{rcharge}), two complex scalar fields of the
theory carry a nonzero charges $\frac{1}{2}$ each. If the Abelian transformation generated by
(\ref{rcharge}) is gauged by a $R$-photon field via a coupling constant $e$,
the correlator we are interested in corresponds to the Coulomb component of the self energy tensor of the photon
to the lowest order in $e$, i.e. $e^2F(q)=\Pi_{00}(0,\vec{q})$, where we consider the static case with $q_0=0$ and $q\equiv|\vec{q}|$.
Its analog in QED is related to the dielectric function in a medium through\cite{LeBellac}
\begin{equation}\label{F}
\epsilon(q)=1+e^2\frac{F(q)}{q^2}.
\end{equation}

\subsection{The one-loop diagrams}

\begin{figure}
\begin{center}
  \includegraphics[width=0.9\linewidth]{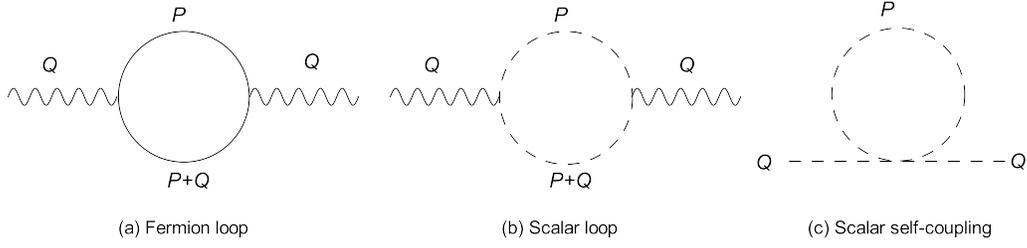}\\
  \caption{One loop self-energy of R-photon. The solid line is for fermion and the dashed line for scalar.}\label{loop}
  \end{center}
\end{figure}

To the zeroth order of the Yang-Mills
coupling, this correlator is given by the one-loop diagram of Fig.1. We have
\begin{equation}
F(q)=(N_c^2-1)\times\Big[2\times\left(\frac{1}{2}\right)^2F_f(q)+2\times\left(\frac{1}{2}\right)^2F_b(q)\Big]
=\frac{N_c^2-1}{2}[F_f(q)+F_b(q)],
\label{F}
\end{equation}
where $F_f(q)$ accounts for the contribution of each Weyl fermion of unit charge and $F_b(q)$ accounts
for the contribution of each complex scalar of unit charge. It follows from the Feynman rule at nonzero
temperature that
\begin{equation}
F_f(q)=T\sum_n\mu^{4-D}\int\frac{d^{D-1}p}{(2\pi)^{D-1}}
{\rm tr}\gamma_4\frac{1}{\displaystyle{\not} P}\gamma_4\frac{1}{\displaystyle{\not} P+\displaystyle{\not} Q}
\label{quarkD}
\end{equation}
with $P=(\nu_n,\vec p)$
and
\begin{equation}
F_b(q)=-4T\sum_n\mu^{4-D}\int\frac{d^{D-1}p}{(2\pi)^{D-1}}
\frac{\omega_n^2}{(\omega_n^2+p^2)[\omega_n^2+(\vec p+\vec q)^2]}
+2T\sum_n\mu^{4-D}\int\frac{d^{D-1}p}{(2\pi)^{D-1}}\frac{1}{\omega_n^2+p^2}.
\label{scalarD}
\end{equation}
where Matsubara frequencies $\nu_n=2\pi T(n+\frac{1}{2})$ and $\omega_n=2\pi Tn$.
The dimensional regularization is applied with $D\to 4^-$ and $\mu$ an energy scale to compensate
the dimensionality. Following the standard procedure in textbooks, we obtain that
\begin{eqnarray}
F_f(q)&=&\frac{q^2}{12\pi^2(4-D)}+\frac{q^2}{24\pi^2}\left(-\ln\frac{q^2}{\mu^2}
+\ln4\pi-\gamma_E+\frac{5}{3}\right)\nonumber\\
&&+\frac{1}{\pi^2}\int_0^\infty dp\frac{p}{e^{\beta p}+1}
\Big[1+\frac{p}{q}\left(1-\frac{q^2}{4p^2}\right)\ln|\frac{2p+q}{2p-q}|\Big]
\label{Ffq}
\end{eqnarray}
and
\begin{equation}
F_b(q)=\frac{q^2}{24\pi^2(4-D)}+\frac{q^2}{48\pi^2}\left(-\ln\frac{q^2}{\mu^2}
+\ln4\pi-\gamma_E+\frac{4}{3}\right)
+\frac{1}{\pi^2}\int_0^\infty dp\frac{p}{e^{\beta p}-1}
\left(\frac{1}{p}+\frac{1}{q}\ln|\frac{2p+q}{2p-q}|\right).
\label{Fbq}
\end{equation}
with $\gamma_E=0.5772...$ the Euler constant.
To explore the analyticity for a complex $q$, it is more convenient to perform
the integrations (\ref{quarkD}) and (\ref{scalarD}) differently. Upon Feynman parametrization of the integrand according to
\begin{equation}
\frac{1}{ab}=\int_0^1dx\frac{1}{[ax+b(1-x)]^2}
\end{equation}
we find that
\begin{equation}
F_f(q)=2 T\sum_n\int_0^1dx\mu^{4-D}\int\frac{d^{D-1}p}{(2\pi)^{D-1}}
\frac{-\nu_n^2+p^2-q^2x(1-x)}{[\nu_n^2+p^2+q^2x(1-x)]^2}
\end{equation}
and
\begin{equation}
F_b(q)=-4T\sum_n\int_0^1dx\mu^{4-D}\frac{d^{D-1}p}{(2\pi)^{D-1}}
\frac{1}{[\omega_n^2+p^2+q^2x(1-x)]^2}
+2T\sum_n\int\mu^{4-D}\frac{d^{D-1}p}{(2\pi)^{D-1}}\frac{1}{\omega_n^2+p^2}.
\end{equation}
It follows that $F_f(q)$ and $F_b(q)$ are analytic on the complex
$q$-plane cut along the imaginary axis. The discontinuities acrossing the cut are given by
\begin{eqnarray}\label{Ffdiscontinuity}
&&\Delta_f(\kappa)\equiv F_f(i\kappa+0^+)-F_f(i\kappa-0^-)=2i{\rm Im}F_f(i\kappa+0^+)\nonumber\\
&=&\frac{T}{\pi^2}\sum_n\int_0^1dx\int_0^\infty
dpp^2[-\nu_n^2+p^2+\kappa^2]
\left( \frac{1}{[\nu_n^2+p^2-\kappa^2x(1-x)+i0^+]^2}-\frac{1}{[\nu_n^2+p^2-\kappa^2x(1-x)-i0^+]^2}\right)\nonumber\\
&=&
2i(N_f+1)T\kappa\Big[\frac{1}{4}-\frac{\pi^2T^2}{3\kappa^2}(4N_f^2+8N_f+3)\Big]
\label{imagf}
\end{eqnarray}
and
\begin{eqnarray}
&&\Delta_b(\kappa)\equiv F_b(i\kappa+0^+)-F_b(i\kappa-0^-)=2i{\rm Im}F_b(i\kappa+0^+) \nonumber\\
&=& -\frac{2T}{\pi^2}\sum_n\omega_n^2\int_0^1dx\int_0^\infty
dpp^2 (\frac{1}{[\omega_n^2+p^2-\kappa^2x(1-x)+i0^+]^2}
-\frac{1}{[\omega_n^2+p^2-\kappa^2x(1-x)-i0^+]^2})\nonumber\\
&=& \frac{i}{3\kappa}\pi^2T^3N_b(N_b+1)(2N_b+1)
\label{imagb}
\end{eqnarray}
where $N_f\ge 0$ and $N_b\ge 0$ are the maximum integers such that $\kappa^2\ge 4\nu_{N_f}^2$ and
$\kappa^2\ge 4\omega_{N_b}^2$. The same analyticity including the discontinuities
can be extracted from (\ref{Ffq}) and (\ref{Fbq}) with a careful treatment of the logarithm
of the integrands. See appendix A for details. We notice that ${\rm Im}F_{f,b}(i\kappa+0^+)>0(<0)$
for $\kappa>0(<0)$, consistent with the identity

\begin{equation}
F_{f,b}(q^*)=F_{f,b}^*(q)
\end{equation}
for an analytic function that is real for real $q$. Substituting (\ref{Ffq}) and (\ref{Fbq}) into
(\ref{F}) we obtain that
\begin{eqnarray}
F(q)&=&\frac{(N_c^2-1)}{16\pi^2}\left\{q^2\ln\frac{1}{ql}
+2\int_0^\infty dp\frac{p}{e^{\beta p}+1}
\Big[1+\frac{p}{q}\left(1-\frac{q^2}{4p^2}\right)\ln|\frac{2p+q}{2p-q}|\Big]\right.\nonumber\\
&&\left.+2\int_0^\infty dp\frac{p}{e^{\beta p}-1}
\left(\frac{1}{p}+\frac{1}{q}\ln|\frac{2p+q}{2p-q}|\right)\right\},
\label{combine}
\end{eqnarray}
where we have introduced a UV cutoff length $l$ via
\begin{equation}
\frac{1}{4-D}=\ln\frac{1}{\mu l}+\frac{1}{2}\left(-\ln4\pi+\gamma_E-\frac{14}{3}\right).
\end{equation}
The discontinuity across the cut along the imaginary axis is given by
\begin{equation}
\Delta(\kappa)\equiv F(i\kappa+0^+)-F(i\kappa-0^+)
=\frac{N_c^2-1}{2}[\Delta_f(\kappa)+\Delta_b(\kappa)].
\end{equation}

\subsection{A multi-loop diagram}

The above analyticity on the momentum plane is generic to a Feynman diagram of a two point Green's
function consisting of an arbitrary
finite number of loops. For a self-energy diagram of $I$ internal lines
and $L$ loops, we have
\begin{equation}
{\cal G}(Q)=\left(\prod_{l=1}^L\sum_{\nu_l}\int\frac{d^{D-1}p_l}{(2\pi)^{D-1}}\right)\frac{N(P,Q)}{\prod_{i=1}^Ik_i^2}
\end{equation}
where $K_i=(\omega_i,\vec k_i)$ is the four momentum of the i-th internal line and $N(P,Q)$ is a polynomial
of loop momenta $P_l=(\nu_l,\vec p_l)$'s and $Q=(q_0,0,0,q)$. Introducing Feynman
parameters, we find that
\begin{eqnarray}
{\cal G}(q)=(I-1)!\int_0^1\prod_{i=1}^I dx_i\delta(1-\sum_{i=1}^Ix_i)
\left(\prod_{l=1}^L\sum_{\nu_l}
\int\frac{d^{D-1}p_l}{(2\pi)^{D-1}}\right)\frac{N(P,Q)}{\left(\sum_{i=1}^I (\omega_i^2+k_i^2)x_i\right)^I}
\end{eqnarray}
where
\begin{equation}
\sum_{i=1}^Ik_i^2x_i=\sum_{ll'}M_{ll'}\vec p_l\cdot\vec p_{l'}+\sum_lu_l\vec p_l\cdot\vec q+vq^2
\end{equation}
with $M_{ll'}$, $u_l$ and $v$ functions of the Feynman parameters $x$'s.
$M_{ll'}$ forms a $L\times L$ symmetric and positive matrix. Upon an orthogonal transformation and a translation,
\begin{equation}
\sum_{i=1}^Ik_i^2x_i=\sum_l\lambda_l\vec p^{\prime 2}+{\cal R}q^2
\end{equation}
with all $\lambda_l$'s positive. Regarding the self-energy part as a DC circuit with
the i-th internal line carrying a resistance of $x_i$, the function ${\cal R}$ equals to
the total resistance between the two terminals of the external lines. We obtain that
\begin{equation}
{\cal G}(q)=(I-1)!\int_0^1\prod_{i=1}^Idx_i\delta(1-\sum_{i=1}^Ix_i)
\left(\prod_{l=1}^L\sum_{\nu_l}\int\frac{d^{D-1}p_l^\prime}{(2\pi)^{D-1}}\right)
\frac{N(P,Q)}{\left(\sum_{i=1}^I\omega_i^2x_i+\sum_l\lambda_l\vec p^{\prime 2}+{\cal R}q^2\right)^I}
\end{equation}
which is analytic throughout the complex $q$-plane except a cut along the
imaginary axis. Note that we did not assume the static limit in the analysis and the result
is not restricted to the Coulomb component of the photon self energy.
While the discontinuity acrossing the cut may be subject to the UV
divergence in general, it should be finite for $N=4$ SYM. On
the other hand, the infrared divergence cannot be removed perturbatively beyond
certain orders, leaving rooms for nonperturbative modifications of the analyticity.

\subsection{A general correspondence}

The location of the singularity is actually dictated by a general
principle. The path integral of a relativistic thermal field theory
at zero chemical potential reads
\begin{equation}
Z={\rm const.}\int[d\Phi]\exp\left(-\int_0^\beta d\tau\int_{-\infty}^{\infty}dx_3\int d^2\vec x_{\perp}
{\cal L}_E[\Phi]\right)
\end{equation}
where $\Phi$ is the collection of the field variables and ${\cal
L}_E$ is the Lagrangian density with a Euclidean time, which is of
$O(4)$ invariance (For a gauge theory, manifest invariance can be implemented in the covariant gauge). The
field theory at nonzero temperature $T=1/\beta$ is formulated on
$R^3$. Switching the role of the Euclidean time $\tau$ and the
spatial coordinate $x_3$, the same path integral describes a Euclidean
field theory at zero temperature formulated on $S^1\times R^2$ with
the same Lagrangian density, described by the partition function
\begin{equation}
{\cal Z}={\rm const.}\int[d\Phi]\exp\left(-\int_{-\infty}^{\infty}d\tau\int_0^\beta dx_3\int d^2\vec x_{\perp}
{\cal L}_E[\Phi]\right)
\end{equation}
The original complex momentum plane associated to the partition
function $Z$ becomes the complex energy plane associated to the
partition function ${\cal Z}$ with the positive imaginary axis of
the former corresponding to the positive real axis of the latter. To
see this mapping clearly, we notice that the spatial coordinate
$x_3$ in $Z$ goes to the Euclidean time $\tau$ of ${\cal Z}$, which
is related to the Minkowski time via $\tau=it$. Accordingly, the
spatial momentum $q$ associated to $Z$ becomes the negative
Euclidean energy $\omega$ associated to ${\cal Z}$ so that
$e^{iqx_3}\to e^{-i\omega\tau}$ and $\omega$ is related to the
Minkowski energy via $\omega=-iE$. It follows then that the spatial
momentum $q$ of $Z$ maps to the Minkowski energy $E$ of ${\cal Z}$
according to $q\to iE$. The unitarity of the field theory specified
by ${\cal Z}$ dictates singularities along the real axis of the
energy plane only, which corresponds to the imaginary axis of the
momentum plane of the original thermal field theory. The spectral
representation of the self energy function implies that its
imaginary part is negative when approaching the cut along the
positive real axis from below and the residues of its poles are all
positive along the positive real axis. The signs of RHS of
(\ref{imagf}) and (\ref{imagb}) as well as the residues of the
asymptotic poles discussed in the next section confirm the above
statement.

%%%%%%%%%%%%%%%%%%%%%%%%%%%%%%%%%%%%%%%%

\section{The AdS/CFT implied correlators}

%%%%%%%%%%%%%%%%%%%%%%%%%%%%%%%%%%%%%%%%
AdS/CFT duality relates the $N=4$ SYM at large $N_c$ and large 't Hooft couping to the following
super-gravity action with a R-photon gauge field
reads\cite{Freedman,Policastro,Huot}
\begin{equation}
S_{\rm sugr}=\frac{N_c^2}{8\pi^2L^3}\int d^5x
\sqrt{g}\left(R+\frac{12}{L^2}\right) +\frac{N_c^2}{64\pi^2L}\int
d^5x\sqrt{g}g^{\mu\rho}g^{\nu\lambda}F_{\mu\nu}F_{\rho\lambda} +\mbox{Gibbons-Hawking term},
\label{sugr}
\end{equation}
where $R$ is the scalar curvature and
\begin{equation}
F_{\mu\nu}=\frac{\partial A_\nu}{\partial x_\mu}-\frac{\partial A_\mu}{\partial x_\nu},
\end{equation}
The metric $g_{\mu\nu}$
is slightly perturbed from the Schwarzschild-AdS$_5$ background, given by
\begin{equation}
ds^2=\frac{L^2}{z^2}\left(fd\tau^2+d\vec x^2+\frac{dz^2}{f}\right)+h_{\mu\nu}dx^\mu dx^\nu
\label{adsblz}
\end{equation}
where $f=1-\pi^4T^4z^4$ with $T$ the temperature of the corresponding field theory.
The action $S_{\rm sugr}$ is a functional of the metric fluctuation
$h_{\mu\nu}(x,z)$ and the R-photon gauge potential $A_\mu(x,z)$ with
$x=(\tau,\vec x)$. Solving the Maxwell equation
\begin{equation}
\frac{\partial}{\partial x^\mu}\sqrt{g}g^{\mu\rho}g^{\nu\lambda}F_{\rho\lambda}=0
\label{maxwell}
\end{equation}
and the linearized Einstein equation
\begin{equation}
R_{\mu\nu}-\frac{1}{2}Rq_{\mu\nu}-\frac{6}{L^2}g_{\mu\nu}=0
\label{einstein}
\end{equation}
subject to the boundary conditions
\begin{equation}
h_{\mu\nu}(x,0)=\bar h_{\mu\nu}(x) \qquad A_\mu(x,0)=\bar A_\mu(x)
\label{boundary}
\end{equation}
under the gauge conditions $A_z(x,z)=h_{\mu z}(x,z)=h_{zz}(x,z)=0$, and
substituting the solutions back to the action, we end up with
\begin{eqnarray}
S_{\rm sugr} &=& S_{\rm sugr}^{(0)}+\frac{1}{2}\int_{u=0}d^4x\int_{u=0}d^4y
[{\cal C}_{\mu\nu}(x-y)\bar A^\mu(x)\bar A^\nu(y)
+\frac{1}{4}{\cal C}_{\mu\nu,\rho\lambda}(x-y)\bar h^{\mu\nu}(x)\bar h^{\rho\lambda}(y)]\nonumber\\
&=& \frac{1}{2}\int\frac{d^4\vec Q}{(2\pi)^4}
[{\cal C}_{\mu\nu}(Q)\bar A^{\mu*}(Q)\bar A^\nu(Q)
+\frac{1}{4}{\cal C}_{\mu\nu,\rho\lambda}(Q)\bar h^{\mu\nu*}(Q)\bar h^{\rho\lambda}(Q)].
\label{generator}
\end{eqnarray}
where the 4D momentum representation has been introduced in the 2nd line of (\ref{generator}).
The coefficients ${\cal C}_{\mu\nu}$ and ${\cal C}_{\mu\nu,\rho\lambda}$ give rise to the
$R$-photon self-energy tensor and the stress tensor correlators.
For the rest of this section, we shall examine the Coulomb and the transverse components of the $R$-photon
self energy and the shear component of the stress tensor correlator in detail. The Matsubara
energy $\omega$ will be set to zero so that the equation of motion of ${\cal C}_{00}$ decouples. The generalization to
a nonzero Euclidean energy will be discussed in the next section. A
UV cutoff is introduced by pulling the 3-brane slightly off the AdS boundary,
i.e. $z=l\to 0^+$. Upon the coordinate transformation, $u=\pi^2T^2z^2$, the metric
(\ref{adsblz}) becomes\cite{Policastro}
\begin{equation}
ds^2=\frac{\pi^2L^2T^2}{u}(f d\tau^2+d\vec
x^2)+\frac{L^2du^2}{4u^2f}+h_{\mu\nu}dx^\mu dx^\nu
\label{adsbl}
\end{equation}
with $f=1-u^2$. The horizon is at $u=1$ and the UV cutoff corresponds to $u=\varepsilon\equiv \pi^2T^2l^2$\cite{footnote1}.
We shall work with this metric below and set the AdS radius $L=1$. The equations of motion
(\ref{maxwell}) and (\ref{einstein}) becomes ordinary differential equations in the 4D momentum
representation and the solution of each equation that satisfies the boundary conditions (\ref{boundary}) and gives
rise to a finite action is unique. The analyticity of these solutions implies that the two
point Green's function we examine {\sl are all meromorphic functions on the complex $q$-plane with infinite number
of poles along the imaginary axis}.

\subsection{The correlator of the $R$-charge density}

It follows from the recipe just stated that the Coulomb component of the $R$-photon self energy
is given by\cite{Policastro}
\begin{equation}
F(q)\equiv {\cal C}_{00}(0,q)=-\frac{N_c^2T^2}{8}\frac{A_0^\prime(\varepsilon|q)}{A_0(\varepsilon|q)},
\label{corr1}
\end{equation}
where $A_0(u|q)$ solves the equation
\begin{equation}
\frac{d^2A_0}{du^2}-\frac{\hat q^2}{u(1-u^2)}A_0=0
\label{A0eq}
\end{equation}
with $\hat q=q/(2\pi T)$. The indexes of this equation at the
canonical singularity $u=1$ are 0 and 1, which give rise to two
linearly independent behaviors as $u\to 0$,
\begin{equation}
A_0(u|q)\simeq 1-u
\label{1st}
\end{equation}
or
\begin{equation}
A_0(u|q)\simeq 1+\frac{1}{2}\hat q^2(1-u)\Big[\ln(1-u)+\frac{1}{2}\Big].
\label{2nd}
\end{equation}
The second solution, eq.(\ref{2nd}) will make the action $S_{\rm sugr}$ diverge and should be
discarded and the eq.(\ref{1st}) generates a power series power series solution
\begin{equation}
A_0(u|q)=(1-u)\sum_{n=0}^\infty a_n(q)(1-u)^n
\label{series1}
\end{equation}
where the coefficients $a_n(q)$ satisfy the recursion relation
\begin{equation}
a_{n+1}(q)=\frac{1}{2(n+1)(n+2)}\left\{[3n(n+1)+\hat
q^2]a_n(q)-n(n-1)a_{n-1}(q)\right\}
\end{equation}
with $a_0(q)=1$. Notice that the coefficients are all polynomials in $q^2$ and is therefore analytic
everywhere on the complex $q$-plane. The indexes at $u=0$ are also 0 and 1 so the series
(\ref{series1}) converges for $0\le 1-u\le 1$ uniformly with respect to $q$.
It follows from the Weierstrass theorem \cite{Weierstrauss}, that the
sum $f(q)\equiv A_0(0|q)$ is an analytic function throughout the
complex $q$-plane and is therefore an entire function.

Next, we consider the derivative $A_0^\prime(u|q)$. The indexes at $u=0$ implies two linearly
independent behaviors there,
\begin{equation}
A_0^{(1)}\sim u\left(1+\frac{1}{2}\hat q^2u\right)
\label{phi1}
\end{equation}
and
\begin{equation}
A_0^{(2)}\sim 1+\hat q^2u\left(\ln u+\frac{1}{2}\right) \label{phi2}
\end{equation}
The solution (\ref{series1}) is a superposition of $A_0^{(1)}$ and $A_0^{(2)}$ and is
dominated by $A_0^{(2)}$ as $u\to 0$, i.e.
\begin{equation}
A_0(u|q)=f(q)[1+\hat q^2u\ln u+O(u)].
\label{smallu}
\end{equation}
and its derivative becomes logarithmically divergent at $u=0$. This
is also consistent with the asymptotic behavior of the coefficients of
(\ref{series1}) at large $n$,
\begin{equation}
a_n(q)\simeq \frac{\hat q^2f(q)}{n(n+1)}
\end{equation}
where the overall factor ${\hat q}^2f(q)$ follows from (\ref{smallu}).
We find then
\begin{equation}
A_0^\prime(\varepsilon|q)=\hat
q^2f(q)\ln\varepsilon+g(q)+O(\varepsilon) \label{series2}
\end{equation}
where
\begin{equation}
g(q)=-1+\sum_{n=1}^\infty\Big[-(n+1)a_n(q)+\frac{1}{n}{\hat
q}^2f(q)\Big ]
\end{equation}
Each coefficient term of the infinite series above is analytic function of $q$ and the series
converges unformily with respect to $q$. The same Weierstrass theorem applied to the
infinite series of $A_0(0|q)$ implies that $g(q)$ is also an entire
function of $q$. Substituting (\ref{series2}) and $A_0(\varepsilon|q)=f(q)+O(\varepsilon\ln\varepsilon)$
into (\ref{corr1}) and dropping the terms that vanish in the limit $\varepsilon\to 0$, we obtain that
\begin{equation}
F(q)\equiv {\cal C}_{00}(0,q)=\frac{N_c^2}{16\pi^2}q^2\ln\frac{1}{\varepsilon}
+\frac{N_c^2T^2}{16\pi^2}\frac{g(q)}{f(q)},
\end{equation}
which is a meromorphic function of $q$.

The poles of $F(q)$ are determined by the zeros of $f(q)$, i.e.
\begin{equation}
A_0(0|q)=0
\label{A0pole}
\end{equation}
It is easy to rule out the zeros off real and imaginary axes, where, ${\rm Im}q^2\neq 0$.
Indeed, taking the imaginary part of the product of $A_0^*(u|q)$ and the equation
(\ref{A0eq}) and integrating $u$ from 0 to 1 yield
\begin{equation}
\left(A_0^*\frac{dA_0}{du}-\frac{dA_0^*}{du}
A_0\right) _{u=0}-{\rm Im}\hat q^2\int_0^1du\frac{A_0^*A_0}{u(1-u^2)}=0
\end{equation}
If there were a complex $q^2$ satisfying (\ref{A0pole}), we would have
\begin{equation}
\int_0^1du\frac{A_0^*A_0}{u(1-u^2)}=0,
\end{equation}
which requires that $A_0(u|q)\equiv 0$. It follows that there cannot
be a nontrivial solution satisfying (\ref{A0pole}) with ${\rm Im}q^2\ne 0$.
Turning to a real $q$, the equation (\ref{A0eq}) implies that the
solution $A_0(u|q)$ is a convex(concave) function of $u$ if
$A_0(u|q)>0(<0)$. Therefore starting with a nonzero $A_0(u|q)$ at $u\ne 0$ there is no way
the solution cannot bent towards
zero at $u=0$ and therefore there is no nontrivial solution satisfying (\ref{A0pole}) in
this case either. The only location for the roots of $A_0(0|q)$ is
the imaginary $q$-axis.

For large and imaginary $q$, i.e. $q=i\kappa$ with $\kappa>>T$,
the WKB approximation of the appendix B yields
\begin{equation}
F(i\kappa)\simeq -\frac{N_c^2}{16\pi^2}\kappa^2\left(\ln\frac{2}{\kappa l}-\gamma_E
+\frac{1}{2}\pi\tan\hat\kappa\delta\right),
\end{equation}
where
\begin{equation}
\delta=\frac{1}{2\sqrt{2\pi}}\Gamma^2\left(\frac{1}{4}\right)
\end{equation}
with $\hat\kappa\equiv\frac{\kappa}{2\pi T}$. Therefore the asymptotic locations of the poles
are given by
\begin{equation}
\kappa\simeq(2n+1)\frac{\pi^2T}{\delta}
\end{equation}
and the number of them is infinite.

\subsection{The self-energy of the transverse $R$-photon}

If follows from the AdS/CFT principle, the self-energy of the transverse $R$-photon
is given by the coefficient of the term quadratic in $\vec A$ of the supergravity
action (\ref{generator}). We have\cite{Policastro}
\begin{equation}
G(q)\equiv {\cal C}_{11}(0,q)={\cal C}_{22}(0,q)=-\frac{N_c^2T^2}{8}\frac{{\cal A}'(\varepsilon|q)}{{\cal A}(\varepsilon|q)}
\end{equation}
with $\varepsilon\to 0$, where ${\cal A}(u|q)$ solves the differential equation
\begin{equation}
-\frac{d}{du}\Big[(1-u^2)\frac{d{\cal A}}{du}\Big]-\frac{\hat q^2}{u}{\cal A}=0.
\label{transverse}
\end{equation}
Both indexes of (\ref{transverse}) at $u=1$ are zero. Only the power series solution leaves the
UV regularized action finite, which reads
\begin{equation}
{\cal A}(u|q)=\sum_{n=0}^\infty b_n(q)(1-u)^n
\label{transpower}
\end{equation}
where the coefficients are given recursively by
\begin{equation}
b_{n+1}(q)=\frac{[\hat q^2+n(3n+1)]b_n(q)-n(n-1)b_{n-1}(q)}{2(n+1)^2}
\end{equation}
with $b_0=1$. The radius of convergence equal to one since none of the
indexes, 0 and 1, at $u=0$ gives rise to divergent behavior there. It follows
from the Weierstrauss theorem of the previous subsection that $G(q)$ is
a meromorphic function of $q$.

Upon a transformation ${\cal A}=\frac{\psi}{\sqrt{1-u^2}}$, the
eq.(\ref{transverse}) becomes a one-dimensional Schroedinger equation
\begin{equation}\label{potential}
-\frac{d^2\psi}{du^2}+V\psi=0
\end{equation}
where the potential
\begin{equation}
V=-\frac{1}{(1-u^2)^2}+\frac{\hat q^2}{u(1-u^2)}.
\end{equation}
and $\psi(u|q)=0$ is required to vanish faster than $\sqrt{1-u}$ as
$u\to 1$. The same argument that rules out the poles of $F(q)$ with a nonzero ${\rm Im}q^2$ can be applied to
the present case but the argument ruling out the poles on the real $q$ axis need to
be modified because of the negative term of the potential. The attractiveness of the potential is maximized at $q=0$.
If there were a nontrivial solution such that $\psi(0|q)=0$ ( i.e.
${\cal A}(0|q)=0)$ for a nonzero real $q$, the zero of ${\cal A}(u|q)$ at $u=0$ would
shift towards $u=1$ when $q^2$ is reduced towards zero.
Consequently, there would be a nontrivial solution at $q=0$, that
vanishes at least once within the domain $u\in(0,1)$. This is,
however, not the case. The exact solution of (\ref{transverse}) for a finite action
is ${\cal A}=1$.
For a large and imaginary $q$, $q=i\kappa$, the WKB approximation yields
\begin{equation}
G(i\kappa)\simeq -\frac{N_c^2}{16\pi^2}\kappa^2\left(\ln\frac{2}{\kappa l}-\gamma_E
-\frac{1}{2}\pi\cot\hat\kappa\delta\right).
\end{equation}

Therefore, the self-energy of a transverse $R$-photon is a meromorphic
function of the momentum $q$ with infinite number of poles along the
imaginary axis.

\subsection{The shear component of the stress tensor correlator}

The shear component of the stress tensor correlator is extracted
from the coefficients of the quadratic term in $h_1{}^2$, ${\cal
C}_1{}^2{}_1{}^2(Q)$ of (\ref{generator}). It follows from the
rotational symmetry about $x_3$-axis that $h_1{}^2=h_2{}^1$ does not
couple with other components of the metric fluctuations and the
quadratic action in $h_1{}^2$ is identical to that of a scalar field
\cite{Policastro}. We have
\begin{equation}
\Pi(q)\equiv
{\cal C}_1{}^2{}_1{}^2(0,q)=-\frac{N_c^2\pi^2T^4}{4}\frac{1-\varepsilon^2}{\varepsilon}
\frac{\phi'(\varepsilon|q)}{\phi(\varepsilon|q)},
\label{shearcomp}
\end{equation}
where $\phi(u|q)$ solves the ordinary differential equation
\begin{equation}
u^2\frac{d}{du}\Big[\frac{1-u^2}{u}\frac{d\phi}{du}\Big]-\hat q^2\phi=0,
\label{shear}
\end{equation}
which is the Fourier transform of the scalar Laplace equation the background metric
(\ref{adsbl}).
Between the two linearly independent solutions, only the power series solution
\begin{equation}
\phi(u|q)=\sum_{n=0}^\infty c_n(q)(1-u)^n
\label{powershear}
\end{equation}
leaves the action (\ref{sugr}) finite upon UV regularization, where the coefficients are given by the
recursion relation
\begin{equation}
c_{n+1}(q)=\frac{[\hat q^2+n(3n-1)]c_n(q)-(n-1)^2c_{n-1}(q)}{2(n+1)^2}
\end{equation}
and are polynomials in $q^2$ again. The power series (\ref{powershear}) is convergent at $u=0$.
Making the transformation $\phi=\sqrt{\frac{u}{1-u^2}}\psi$, the equation (\ref{shear}) can be converted
into an one-dimensional Schroedinger equation of zero energy in the potential
\begin{equation}
V=\frac{3}{4u^2}-\frac{u^2}{(1-u^2)^2}+\frac{\hat q^2}{u(1-u^2)}
\end{equation}
which is the least repulsive at $q=0$. The exact solutions of
(\ref{shear}) at $q=0$ is $\phi(u|q)=1$. For $q=i\kappa$ with
$\kappa>>T$, we find that
\begin{equation}
\Pi(i\kappa)\simeq -\frac{N_c^2}{16\pi^2l^2
}\kappa^2-\frac{N_c^2}{32\pi^2}\kappa^4
\left(\ln\frac{2}{\kappa l}-\gamma_E
+\frac{1}{2}\pi\tan\hat\kappa\delta\right).
\end{equation}
Therefore the arguments applied in the last subsection for the transverse
$R$ photon self-energy implies that $\Pi(q)$ is a meromorphic function with
an infinite number of poles along the imaginary axis of the $q$-plane, like
the other two cases.

%%%%%%%%%%%%%%%%%%%%%%%%%%%%%%%%%%%%%%%%

\section{Discussions}
Let us recaptulate what we have done in previous sections. We
started with the example of the $R$-charge density correlator of the
$N=4$ SYM in the static limit and worked out the its analyticity on
the complex momentum plane. Through the correspondence between a
relativistic field theory at a nonzero temperature formulated in
$R^3$ and another relativistic field theory at zero temperature
formulated in $S^1\times R^2 $, we have found the universality
regarding the location of the singularities on the momentum plane of
a two point Green's function. Perturbatively, the singularities form
a branch cut along the imaginary axis, reflecting the continuum of
asymptotic states of the field theory on $S^1\times R^2$. Our
conclusion in this regard, however, is not limited to the $N=4$ SYM
being considered and is also valid with a nonzero Euclidean energy.

A by product of our analysis is to rule out the Yukawa oscillations of certain static potential
in a hot medium at zero chemical potential, as was suggested in the literature\cite{liuh,ChengfuMu}. Taking the Coulomb potential as an example,
which takes the form
\begin{equation}
U(r)=\frac{e^2}{4\pi^2r}{\rm Im}\int_{-\infty}^\infty dq\frac{qe^{iqr}}{q^2+e^2F(q)}.
\end{equation}
with $e^2<<1$. Here both $e$ and $F(q)$ refer to renormalized
quantities which amounts to replace the cutoff length $l$ of
(\ref{combine}) with a renormalization length scale comparable to
$1/T$. A complex pole at $q=q_R+iq_I$ with $q_R\ne 0$ will make the
function $U(r)$ oscillates in $r$. Because $F(q)$ grows like
$\ln|q|$ for large $|q|>>T$, there is a tachyonic pole $q=K_T$ of
the integrand on the real axis with $\ln K_T\sim \frac{1}{e^2}$. The
oscillation it causes can be eliminated easily by smearing the
source charge of the potential. To explore the poles with $|q|\sim
T$, we write $d(q)\equiv q^2+e^2F(q)$ and look for its roots within
a circle $|q|=K$ with $T<<K<<K_T$, given by the intersection of the
lines ${\rm Re}\ d(q)=0$ and ${\rm Im}\ d(q)=0$. Because of the
symmetry of $F(q)$, we may restrict our searching scope within the
first quadrant. A line of ${\rm Im}\ d(q)=0$ cannot cross the cut
along the imaginary axis since ${\rm Im}\ d(q)\ne 0$ there. It
cannot form a loop within the region where $d(q)$ is analytic. So it
has to cross the circle just introduced. On the circle $d(q)\simeq
q^2$ and we need only to examine the neighborhood where the circle
cuts the real and imaginary axes. The real axis itself is a line
where ${\rm Im}d(q)=0$ and $d(q)>0$ for $|q|\leq K$ There is another
one near the imaginary axis which can be located by treating ${\rm
Im}F(q)$ perturbatively. The positivity of ${\rm Im}F(q)$ for $q$
approaching the positive imaginary axis from right pushes the line
${\rm Im}d(q)=0$ off the first quadrant. Therefore there is no
complex zero of $d(q)$ that gives rise to the Yukawa
oscillation\cite{liuh,ChengfuMu,Alonso1,Alonso2}.

Next we moved to the strong coupling limit where exact expression of the two point Green's function
exists for $N=4$ SYM at large $N_c$ and large 't Hooft coupling through AdS/CFT duality. There
we examined the $R$-charge density correlator, the self-energy of a transverse $R$-photon and the correlator of the
shear component of the stress tensor, by solving the equation of motion of their gravity dual.
In the static limit it follows from the series representation and the Weierstrass theorem that these
functions are all analytic on the momentum plane except the imaginary axis. But instead of a branch cut
running there, we have found infinite number of poles, which signify a bona fide nonperturbative effect. More impressive
is to view the energy plane of the $N=4$ SYM in $S^1\times R^2$ at zero temperature, show in Fig.2.
The continuum of asymptotic states is strongly distorted into a set of bound states and their asymptotic distribution
$E_n\simeq \frac{2n\pi^2T}{\delta}$ for $n>>1$ resembles that of the mass spectrum of
AdS/QCD with a hard wall\cite{hardwall}. Here is a possible mechanism: The composite states excited by the R-charge
current operators or the stress tensors are all color singlet. The continuous energy spectrum
would correspond to a composite wave function whose constituents can be separated indefinitely. While this
is the case for a cluster of free particles but may be prohibited when Yang-Mills interaction is turned on. If the extension
of the wave function in $R^2$ dimensions is much longer than the radius of $S^1$, the
binding dynamics is essentially 2D and a Yang-Mills theory is expected to confine with two spatial dimensions \cite{nair}.
Also the color singlet intermediate states cannot decay into separate color singlet objects since the corresponding
Feynman diagrams are suppressed at large $N_c$\cite{nair_1}.

\begin{figure}
\begin{center}
  \includegraphics[width=\linewidth]{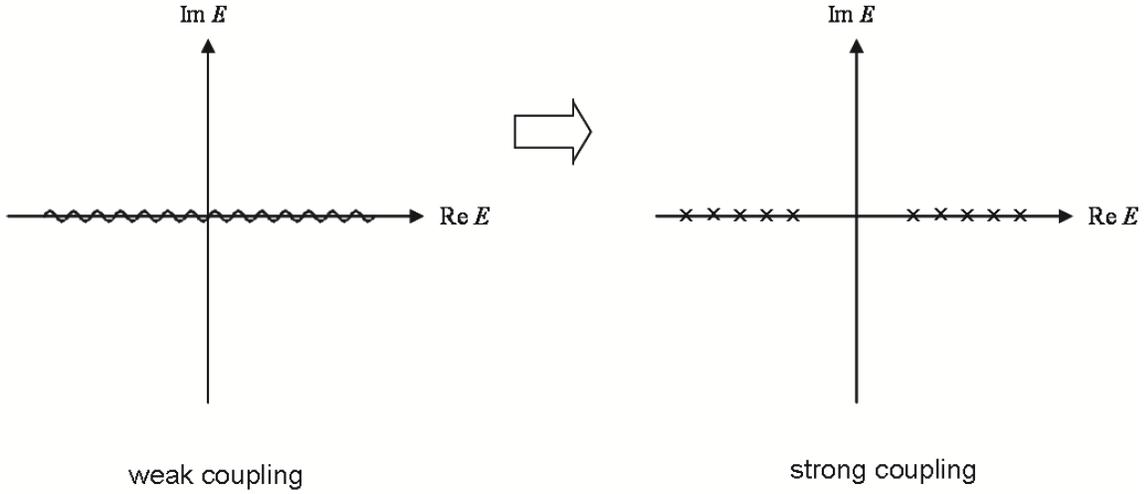}\\
  \caption{The energy plane of N=4 SYM on $S^1\times R^2$ at zero temperature. The left is for weak coupling with the wiggly line representing branch cut and the right is for strong coupling with the crosses representing poles.}\label{branchcut}
  \end{center}
\end{figure}

The generalization to a nonzero Euclidean energy is straightforward for the transverse $R$-photon self energy
and the shear component of the stress tensor correlator, since their equation of motion remains uncoupled.
For the transverse $R$-photon of a Euclidean energy $\omega$, the EOM (\ref{transverse}) becomes
\begin{equation}
-\frac{d}{du}\Big[(1-u^2)\frac{d{\cal A}}{du}\Big]-\left(\hat q^2+\frac{\hat\omega^2}{1-u^2}\right)
\frac{{\cal A}}{u}=0
\label{eom}
\end{equation}
with $\hat\omega=\frac{\omega}{2\pi T}$ and solution (\ref{transpower}) takes the form
\begin{equation}
{\cal A}(u|Q)=(1+u)^\alpha(1-u)^\beta\sum_{n=0}^\infty b_n(Q)(1-u)^n
\end{equation}
with $\alpha=-\frac{i}{2}\omega$ and $\beta=\frac{1}{2}|\omega|$. The recursion formula of the
coefficients reads
\begin{eqnarray}
b_{n+1}(Q)&=&\frac{1}{2(n+1)(n+1+2\beta)}\left\{[\hat\omega^2+\hat q^2+\alpha+\beta+2\alpha\beta
+n(3n+1+2\alpha+6\beta)]b_n(Q)\right.\nonumber\\
&&\left.-[(n-1)(n+2\alpha+2\beta)+\alpha+\beta+2\alpha\beta]b_{n-1}(Q)\right\}
\end{eqnarray}
with $b_0(Q)=1$ and all coefficients remain polynomials of $q^2$. Furthermore, the indexes
at $u=0$ remains 0 and 1. Upon transforming eq.(\ref{eom}) into a Schroedinger equation of the
form (\ref{potential}), we have the potential
\begin{equation}
V=-\frac{1}{(1-u^2)^2}+\frac{\hat q^2}{u(1-u^2)}+\frac{\hat\omega^2}{u(1-u^2)^2},
\end{equation}
which is more repulsive than the static case. Therefore
the same analysis of the last section leads to the conclusion that the self-energy
is a meromorphic function of a complex momentum at $\omega\ne 0$ and has infinite number of
poles along the imaginary axis. Likewise is the shear component of the stress tensor
correlator at $\omega\ne 0$. Notice that the coefficients of the power series depends on
$|\omega|$ because only the vanishing solution at the horizon is selected. This also reflects
the cut along the real axis of the energy plane.

%%%%%%%%%%%%%%%%%%%%%%%%%%%%%%%%%%%%%%%%

%%%%%%%%%%%%%%%%%%%%%%%%%%%%%%%%%%%%%%%%

\section*{Acknowledgments}

We would like to extend our gratitude to Mei Huang, V. P. Nair, Zi-wei Lin and Peng-fei Zhuang
for helpful discussions. The work of D. F. H. and H. C. R.
is supported in part by NSFC under grant Nos. 10975060, 10735040. The work of Hui Liu is supported in part by NSFC under grant No. 10947002.
%The work of D.F. H. is also supported in part by Educational Committee under
%grants NCET-05-0675 project No. IRT0624.

%\acknowledgments

%%%%%%%%%%%%%%%%%%%%%%%%%%%%%%%%%%%%%%%%

\appendix
\section{Analyticity of self-energy in complex $q$-plane }

A direct analytical continuation to complex $q$-plane of the
integral representation of the one-loop self energy functions
$F_f(q)$ and $F_b(q)$ of Eqs.(\ref{Ffq}) and (\ref{Fbq}) is obscured
by the logarithm of the integrand. In this appendix, we shall show
how this is carried out for $F_f(q)$ in details.

Let us break $F_f(q)$ into two the contributions from the vacuum
polarization and the thermal fluctuations,
\begin{equation}
F_f(q)=F_{\rm vac}(q)+F_{\rm med}
\end{equation}
with
\begin{equation}
F_{\rm
vac}(q)=\frac{q^2}{12\pi^2(4-D)}+\frac{q^2}{24\pi^2}\left(-\ln\frac{q^2}{\mu^2}
+\ln4\pi-\gamma_E+\frac{5}{3}\right) \label{vac}
\end{equation}
and
\begin{equation}
F_{\rm med}(q)=\frac{1}{\pi^2}\int_0^\infty dp\frac{p}{e^{\beta
p}+1}
\Big[1+\frac{p}{q}\left(1-\frac{q^2}{4p^2}\right)\ln|\frac{2p+q}{2p-q}|\Big].
\label{med}
\end{equation}
We draw the branch cut of the logarithm from $-q/2$ to $q/2$ on
$p-plane$. To accomplish the continuation, one could first implement
the absolute value under the logarithm with two integral paths, one
above and one below the cut for a real $q$ and then deform the paths
accordingly as $q$ becomes complex, i.e.

\begin{figure}
\begin{center}
  \includegraphics[width=\linewidth]{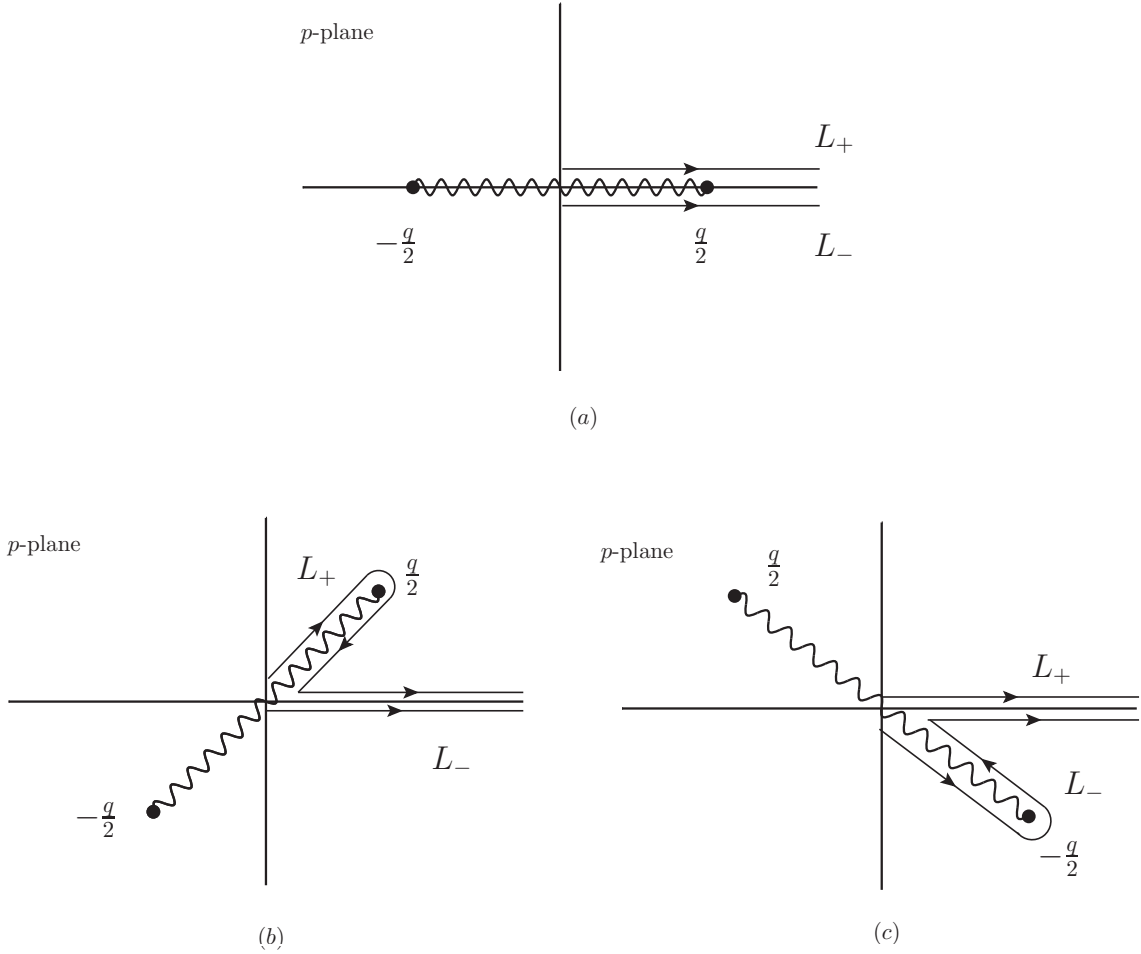}\\
  \caption{Analytical continuation of complex $q$. The two integral paths in (a) are deformed as (b) and (c) for real
$q>0$ and $q<0$ respectively.}\label{cut}
  \end{center}
\end{figure}

\begin{eqnarray}
F_{med}(q)&=&\frac{1}{\pi^2}\int_0^\infty dp\frac{p}{e^{\beta p}+1}
\left(1+\frac{4p^2-q^2}{4pq}\ln\left|\frac{2p+q}{2p-q}\right|\right) \\
&=&\frac{1}{2\pi^2}\left(\int_{L_+}
+\int_{L_-}\right)dp\frac{p}{e^{\beta p}+1}
\left[1+\frac{4p^2-q^2}{4pq}\ln\left(\frac{2p+q}{2p-q}\right)\right]
\end{eqnarray}
where the two integral path $L_+$ and $L_-$ are defined slightly
above and below the positive real axis as shown in Fig.\ref{cut}(a).
The sections with $p>q/2$ of $L_\pm$ make no difference to the
integral. To find out the discontinuity of $F_{med}$ across the
imaginary axis, one has to calculate
$F_{med}(i\kappa+0^+)-F_{med}(i\kappa-0^+)$ where $\kappa$ is an
arbitrary real positive number and $0^+$ the positive infinitesimal.
The case of negative $\kappa$ can be obtained by symmetry. Now we
are ready to make the analytical continuation for both positive and
negative real $q$.

\begin{itemize}
  \item For real $q>0\longrightarrow q=q_R+iq_I$ where $q_R>0, q_I>0$.

  We rotate the branch cut of $p$-plane counter-clockwise and deform the $L_+$ path along the cut as shown in
Fig.\ref{cut}(b). Then $F_{med}$ contains contribution both from the
path along positive real axis and that around the cut, i.e.,
  \begin{equation}
    F_{med}(q)=F_1^+(q)+F_2^+(q)
  \end{equation}
  \begin{eqnarray}
    F_1^+(q)&=&\frac{1}{\pi^2}\int_0^\infty dp\frac{p}{e^{\beta p}+1}
\left[1+\frac{4p^2-q^2}{4pq}\ln\left(\frac{2p+q}{2p-q}\right)\right]\\
    F_2^+(q)&=&-\frac{i}{4\pi q}\int_0^{\frac{q}{2}} dp\frac{4p^2-q^2}{e^{\beta p}+1}
  \end{eqnarray}

  \item For real $q<0\longrightarrow q=q_R+iq_I$ where $q_R<0, q_I>0$.

  Similarly, we rotate the cut clockwise to realize the continuation as shown in Fig.\ref{cut}(c).
Notice that now real $q<0$ and $F_{med}$ is decomposed as
  \begin{equation}
    F_{med}(q)=F_1^-(q)+F_2^-(q)
  \end{equation}
  \begin{eqnarray}
    F_1^-(q)&=&\frac{1}{\pi^2}\int_0^\infty dp\frac{p}{e^{\beta p}+1}
\left[1+\frac{4p^2-q^2}{4pq}\ln\left(\frac{2p+q}{2p-q}\right)\right]\\
    F_2^-(q)&=&-\frac{i}{4\pi q}\int_0^{-\frac{q}{2}} dp\frac{4p^2-q^2}{e^{\beta p}+1}
  \end{eqnarray}
\end{itemize}

Let $q=i\kappa\pm0^+$ and consider that
$F_1^+(i\kappa)-F_1^-(i\kappa)=0$, then the discontinuity across the
imaginary $q$ axis reads
\begin{equation}
  F_{med}(i\kappa+0^+)-F_{med}(i\kappa-0^+)=F_2^+(i\kappa+0^+)-F_2^-(i\kappa-0^+)
\end{equation}
Rewriting
\begin{eqnarray}
F_2^+(i\kappa+0^+)=-\frac{1}{4\kappa\pi}\int_{0^+}^{\frac{i}{2}\kappa+0^+}
dp\frac{4p^2+\kappa^2}{e^{\beta p}+1}
=-\frac{1}{4\pi\kappa}\int_{0^-}^{\frac{i}{2}\kappa-0^+}
dp\frac{4p^2+\kappa^2}{e^{\beta p}+1}+\triangle
\end{eqnarray}
where
\begin{eqnarray}
\triangle&\equiv&-\frac{1}{4\pi\kappa}\left(\int_{0^+}^{\frac{i}{2}\kappa+0^+}
-\int_{0^-}^{\frac{i}{2}\kappa-0^+}\right) dp\frac{4p^2+\kappa^2}{e^{\beta p}+1}\nonumber\\
&=&-\frac{i}{2\kappa}\sum_{\tiny 0<\nu_n<\kappa/2}\mbox{Res}\left(\frac{4p^2+\kappa^2}{e^{\beta p}+1};p=i\nu_n\right)\\
&=&2iT\kappa\sum_{\nu_n>0}\left(\frac{1}{4}-\frac{\nu_n^2}{\kappa^2}\right)\\
&=&2i(N_f+1)T\kappa\left[\frac{1}{4}-\frac{\pi^2T^2}{3\kappa^2}(4N_f^2+8N_f+3)\right],
\end{eqnarray}
where $N_f\ge 0$ is the maximum integer such that $\kappa^2\ge
4\nu^2_{N_f}$. Combining (\ref{vac}) with (\ref{med}), one finds
that all terms except $\triangle$ cancel out, leading the
discontinuity of Eq.(\ref{Ffq}) to
\begin{eqnarray}
F_f(i\kappa+0^+)-F_f(i\kappa-0^+)&=&-\frac{(i\kappa+0^+)^2}{24\pi^2}\ln\frac{(i\kappa+0^+)^2}{\mu^2}
+\frac{(i\kappa-0^+)^2}{24\pi^2}\ln\frac{(i\kappa-0^+)^2}{\mu^2}\nonumber\\
&&-\frac{1}{4\pi\kappa}\left(\int_{0^-}^{\frac{i}{2}\kappa-0^+}-\int_{0^+}^{-\frac{i}{2}\kappa+0^+}\right)
dp\frac{4p^2+\kappa^2}{e^{\beta p}+1}+\triangle\\
&=&\frac{i\kappa^2}{12\pi}-\frac{1}{4\pi\kappa}\int^{\frac{i}{2}\kappa-0^+}_{0^-}dp(4p^2+\kappa^2)+\triangle\\
&=&\triangle
\end{eqnarray}
in agreement with Eq.(\ref{Ffdiscontinuity}).

\section{WKB approximation for a large and imaginary $q$}

For imaginary momentum,we set $\hat{q}=i\hat{\kappa}$, then
Eq.(\ref{A0eq}) becomes
\begin{equation}\label{originaleom}
\frac{d^2A_0}{du^2}+\frac{\hat \kappa^2}{u(1-u^2)}A_0=0
\end{equation}
We are going to find out $A_0(0|\kappa)$ by solving this equation
for $\hat\kappa>>1$ with WKB approximation.

Near the $u=0$ and $u=1$ regions, the equation of motion can be
rewritten as
\begin{eqnarray}
&&\frac{d^2A_0}{du^2}+\frac{\hat \kappa^2}{u}A_0=0\\
&&\frac{d^2A_0}{du^2}+\frac{\hat \kappa^2}{2(1-u)}A_0=0
\end{eqnarray}
respectively, which can be reduced to the standard Bessel equations
with the solutions
\begin{eqnarray}
  &&A_0(u\rightarrow 0|\kappa)=\sqrt{u}[C_1 H_1^{(1)}(2\hat\kappa \sqrt u)+C_2 H^{(2)}_1(2\hat\kappa \sqrt u)]
\label{u=0sol}\label{hankel}\\
  &&A_0(u\rightarrow 1|\kappa)=\sqrt{1-u}J_1(\hat\kappa\sqrt{2(1-u)})\label{u=1sol}
\end{eqnarray}
where $H_1^{(1)}$ and $H_1^{(2)}$ are the first and second kind of
Hankel functions respectively, and $J_1$ is the Bessel function.
$C_1$ and $C_2$ are the combination coefficients. Notice that in the
solution near $u=1$, considering the vanishing contribution from the
horizon, one must reserve only one independent solution.

When $\hat\kappa\sqrt{u(1-u^2}>>1$, one could implement the WKB
approximation and obtain
\begin{equation}\label{WKB}
A_0^{WKB}=[u(1-u^2)]^\frac{1}{4}\left[C_+\exp\left(i\hat\kappa\int_u^1\frac{dx}{\sqrt{x(1-x^2)}}\right)
+C_-\exp\left(-i\hat\kappa\int_u^1\frac{dx}{\sqrt{x(1-x^2)}}\right)\right]
\end{equation}
In the matching region of validity where $1-u<<1$ but
$\hat\kappa\sqrt{2(1-u)}>>1$, where both (\ref{u=1sol}) and
(\ref{WKB}) work as the solution of (\ref{originaleom}), one could
expand(\ref{u=1sol}) as
\begin{eqnarray}
A_0(u|\kappa)&\approx&\frac{2^\frac{1}{4}}{\sqrt{\pi\hat\kappa}}(1-u)^\frac{1}{4}
\sin\left[\hat\kappa\sqrt{2(1-u)}-\frac{\pi}{4}\right]\\
&\approx&\frac{2^{-\frac{3}{4}}}{i\sqrt{\pi\hat\kappa}}(1-u)^\frac{1}{4}
\left[\exp\left(i\hat\kappa\sqrt{2(1-u)}-i\frac{\pi}{4}\right)
-\exp\left(-i\hat\kappa\sqrt{2(1-u)}+i\frac{\pi}{4}\right)\right].\label{expan}
\end{eqnarray}
and rewrite the WKB solution (\ref{WKB}) as
\begin{equation}\label{WKBexpan}
A_0^{WKB}(u|\kappa)=2^\frac{1}{4}(1-u)^\frac{1}{4}\left[C_+\exp(i\hat\kappa\sqrt{2(1-u)}
+C_-\exp(-i\hat\kappa\sqrt{2(1-u)}))\right].
\end{equation}
Then one obtains
\begin{equation}
C_+=\frac{1}{2i\sqrt{\pi\hat\kappa}}e^{-i\frac{\pi}{4}}, C_-=C_+^*
\end{equation}
and the WKB solution (\ref{WKB}) becomes
\begin{equation}
A_0^{WKB}=\frac{u^{\frac{1}{4}}(1-u^2)^{\frac{1}{4}}}{2i\sqrt{\pi\hat\kappa}}
\left[\exp\left(i\hat\kappa\int^1_u
\frac{dx}{\sqrt{x(1-x^2)}}-i\frac{\pi}{4}\right)
-\exp\left(-i\hat\kappa\int^1_u
\frac{dx}{\sqrt{x(1-x^2)}}+i\frac{\pi}{4}\right)\right].
\end{equation}
This expression remains efficient in the region where $u<<1$ but
$2\hat\kappa\sqrt u>>1$ and we have
\begin{equation}\label{wkb1}
A_0^{WKB}(u|\kappa)\approx\frac{u^\frac{1}{4}}{2i\sqrt{\pi\hat\kappa}}
\left[\exp\left(i\hat\kappa\delta-i\frac{\pi}{4}-2i\hat\kappa\sqrt{u}\right)\right]
-\left[\exp\left(-i\hat\kappa\delta+i\frac{\pi}{4}+2i\hat\kappa\sqrt{u}\right)\right]
\end{equation}
where
$\delta\equiv\frac{1}{2\sqrt{2\pi}}\Gamma^2\left(\frac{1}{4}\right)$.
In Eq.(\ref{wkb1}) we have used the approximation
\begin{equation}
\int_u^1\frac{dx}{\sqrt{x(1-x^2)}}=\int_0^1\frac{dx}{\sqrt{x(1-x^2)}}-\int_0^u\frac{dx}{\sqrt{x(1-x^2)}}\approx\delta-2\sqrt
u
\end{equation}
when $u<<1$. We match the WKB solution to the analytic form of
Eq.(\ref{hankel}) in $u<<1$, and the solution all the way to $u=0$
is
\begin{equation}
  A_0^{WKB}(u|\kappa)=\frac{\sqrt{u}}{2i}
\left[e^{-i\hat\kappa\delta}H_1^{(1)}(2\hat\kappa \sqrt
u)-e^{i\hat\kappa\delta}H_1^{(2)}(2\hat\kappa \sqrt u)\right]
=A_0(u\rightarrow 0|\kappa).
\end{equation}
In the calculation we used the approximation formulae of Hankel
function when $2\hat\kappa\sqrt u>>1$
\begin{eqnarray}
H_1^{(1)}(2\hat\kappa\sqrt
u)&\approx&-\frac{u^{-\frac{1}{4}}}{\sqrt{\pi \hat\kappa}}
\exp\left[i\left(2\hat\kappa\sqrt u +\frac{\pi}{4}\right)\right],\\
H_1^{(2)}(2\hat\kappa\sqrt
u)&\approx&-\frac{u^{-\frac{1}{4}}}{\sqrt{\pi \hat\kappa}}
\exp\left[-i\left(2\hat\kappa\sqrt u +\frac{\pi}{4}\right)\right].
\end{eqnarray}

According to Eq.(\ref{corr1}), the correlator of R-charge density
with imaginary momentum reads
\begin{eqnarray}
F(i\kappa)&=&-\frac{N_c^2 T^2\hat\kappa}{8\sqrt
\varepsilon}\frac{e^{-i\hat\kappa\delta}H_0^{(1)}(2\hat\kappa \sqrt
\varepsilon) -e^{i\hat\kappa\delta}H_0^{(2)}(2\hat\kappa \sqrt
\varepsilon)} {e^{-i\hat\kappa\delta}H_1^{(1)}(2\hat\kappa \sqrt
\varepsilon)
-e^{i\hat\kappa\delta}H_1^{(2)}(2\hat\kappa \sqrt \varepsilon)}\\
&\approx&\frac{N_c^2}{16\pi^2}\left[-\kappa^2\left(\ln\frac{2}{\kappa
l}-\gamma_E\right)
-\frac{1}{2}\pi\kappa^2\tan(\hat\kappa\delta)\right]
\end{eqnarray}
where $l=\varepsilon/(\pi^2T^2)$.

%%%%%%%%%%%%%%%%%%%%%%%%%%%%%%%%%%%%%%%%


\begin{thebibliography}{99}
\bibitem{footnote1} The UV cutoff can also be implemented via a dimensional regulariztion, which amounts to write
$\sqrt{g}=\pi T\sqrt{\frac{f}{u}}\left(\frac{\pi T}{\sqrt{u}}\right)^{D-1}\frac{1}{2u\sqrt{f}}
=\frac{1}{2}(\pi T)^Du^{-\frac{D}{2}-1}$ with $D\to 4^-$.
\end{thebibliography}
\end{document}